\documentclass[prc,showpacs,floatfix]{revtex4}
\usepackage{psfig}

\newcommand{\bra}[1]{\mbox{$\langle #1|$}}
\newcommand{\ket}[1]{\mbox{$|#1\rangle$}}

\begin{document}

\title
{\hfill{\small {\bf MKPH-T-04-01}}\\
{\bf Meson retardation in deuteron electrodisintegration}\footnote[2]{Supported by the Deutsche Forschungsgemeinschaft (SFB
443).}
}
\author{M.\ Schwamb and  H.\ Arenh\"ovel}
\affiliation{Institut f\"ur Kernphysik, Johannes
Gutenberg-Universit\"at, D-55099 Mainz, Germany} 
\date{\today}                                     

\begin{abstract}
\noindent
The effect of meson retardation in $NN$-interaction and exchange
currents on deuteron electrodisintegration is studied in a coupled
channel approach including $NN$-, $N \Delta$- and $\pi d$-channels.
It is shown that the influence of retardation depends on the energy
regime: Whereas below $\pi$-threshold calculations with static and
retarded operators yield almost identical results, they differ
significantly in the $\Delta$-region. Especially, the longitudinal
and the longitudinal-transverse interference
structure functions are strongly affected. 

\end{abstract}

\maketitle

\section{Introduction}
\label{intro}

Recently we have constructed a realistic $NN$-interaction model and
corresponding electromagnetic (e.m.) exchange currents based on a coupled
channel approach with meson, nucleon and $\Delta$-degrees of freedom
and have applied it to $NN$-scattering and deuteron
photodisintegration~\cite{ScA98,ScA01a,ScA01b}. It contains complete
meson retardation in both the $NN$-interaction as well as in the
e.m.\ pionic meson exchange currents ($\pi$-MEC). In addition off-shell
contributions to the e.m.\ one-body current were considered in
subsequent work~\cite{ScA01c}. As was shown in~\cite{ScA01b}, the
influence of retardation is large above pion threshold and only its
incorporation leads to a satisfactory theoretical description of
deuteron photodisintegration in the $\Delta$-region. On the other hand, e.m.\
off-shell effects turned out to be quite small~\cite{ScA01c}.

Due to the possibility of an independent variation of energy and momentum
transfer in the space-like region and the existence of a longitudinal
polarization of the exchanged virtual photon, it is expected that
electrodisintegration of the deuteron will provide additional insights
into the hadronic properties of the $NN$-system. For example, it
allows the study of the $NN$-interaction in the short-range regime 
independent of the excitation energy in contrast to
photodisintegration. Moreover, in quasifree kinematics this
reaction is an important tool for the exctraction of nucleon
properties like, e.g.\ the electric form factor of the
neutron~\cite{Her99,Ost99,Pas99}. For this reason, we have extended
our approach to electrodisintegration. In the next section,
a brief survey of our model is given. Then, in Sect.~\ref{result},
results are presented and compared to experimental data 
for some selected kinematics and, finally, some conclusions are drawn.

\section{Theoretical framework}
\label{model}
The model Hilbert space consists of three basic configurations, i.e.\
two nucleons ($NN$), one nucleon and one delta ($N\Delta$), and two
nucleons and one meson (e.g.\ $NN\pi$). The basic hadronic interactions
are determined by baryon-baryon-meson vertices, where we include as
mesons $\pi,\, \rho,\, \sigma,\, \delta,\, \omega$, and $\eta$. 
Inserting the vertices into the appropriate Lippmann-Schwinger
equation, one obtains after some straightforward algebra~\cite{ScA01a} 
the desired retarded one-boson-exchange operators describing the
interactions $NN  \leftrightarrow NN$, $NN \leftrightarrow N \Delta$, and
$N \Delta \leftrightarrow N \Delta$. In our explicit realization, we use
for the parametrization of the retarded $NN$-interaction the Elster
potential~\cite{ElF88}, which includes in addition one-pion
loop diagrams as nucleon self energy contributions in order to fulfil
unitarity above pion threshold. For this reason, one has to
distinguish between bare and physical nucleons. For details we refer
to~\cite{ScA01a}. For the interactions $NN \leftrightarrow N \Delta$
 and $N \Delta \leftrightarrow N\Delta$, we take besides retarded
pion exchange in addition only static $\rho$-exchange into
account. Moreover, the interaction of two nucleons in the channel with
deuteron quantum numbers in the presence of a spectator pion, 
called $\pi d$-channel, is also considered. By a suitable 
box renormalization~\cite{GrS82}, an approximate phase equivalence
between the Elster potential, which does not include $\Delta$-d.o.f., and
our coupled channel approach below 
pion threshold is obtained. The hadronic $\Delta$-parameters
are fitted to the $P_{33}$-channel of pion-nucleon scattering and
the $^1D_2$-channel in $NN$-scattering.
For the sake of comparison with the conventional approach with static
interactions, we have constructed also a static version of our model
interaction based on the Bonn OBEPR-potential~\cite{MaH87a} where in
accordance with the usual treatment the pion-nucleon loop diagrams
have been neglected. For the results presented below, we have used 
two of the various model versions discussed in~\cite{ScA01a}, namely 
``CC(ret,$\pi$,$\rho$,0)" and  ``CC(stat,$\pi$,$\rho$,0)" for the
retarded and static calculations, respectively.

The basic e.m.\ interaction of the model is described in
detail in~\cite{ScA01b,ScA01c} and consists of baryon and meson
one-body currents as well as Kroll-Rudermann (contact) and vertex 
contributions. These currents are, together with the $\pi NN$-vertex,
the basic building blocks of the corresponding effective current
operators for two baryons consisting of one- and two-body
contributions. The matrix element of the effective nucleon one-body
operator between plane-wave states of two bare nucleons ${\bar N}$ is
given in~\cite{ScA01c} for real photons. Its extension to virtual
photons with four momentum $q^\mu=(\omega,\vec q\,)$ reads as follows 
\begin{eqnarray}
 \bra{ \vec{p}_1^{\,\prime},\,  \vec{p}_2^{\,\prime}\,} 
J_{eff}^{N[1]\, \mu}(z,\omega,\vec{q}\,)
 \ket{ \vec{p}_1,\,\vec{p}_2\,}  &=& 
\bra{ \vec{p}_1^{\,\prime},\, \vec{p}_2^{\,\prime}\,} 
 \frac{{\widehat R}(z)}{{\widehat R}(z^{os})}
 j_{real}^{N[1]\, \mu}(\omega,\vec{q}\,)
 \frac{{\widehat R}(z-\omega)}{{\widehat R}(z^{os}-\omega)}
 \ket{ \vec{p}_1,\,\vec{p}_2\,}   \nonumber \\  
 && +\bra{ \vec{p}_1^{\,\prime},\,\vec{p}_2^{\,\prime}\,} 
 {\widehat R}(z) {\cal J}_{loop,\, sub}^{\mu}(z,\omega,\vec{q}\,) 
 {\widehat R}(z-\omega)
 \ket{ \vec{p}_1,\,\vec{p}_2\,}\,, \label{3_final}
\end{eqnarray}
where $j_{real}^{N[1]\,\mu}(\omega,\vec{q}\,)$ denotes the usual
nucleon one-body four-current including leading order relativistic
contributions like Darwin-Foldy and spin-orbit. They contain e.m.\
Sachs form factors using the dipole parametrization. Explicit formulas
can be found, for example, in~\cite{WiB93}. For simplicity,
the electric form factor of the neutron is neglected. Furthermore, 
$z= W + i \epsilon$, and $z^{os} = \sqrt{M_N^2+
(p_1')^2} + \sqrt{M_N^2 + (p_2')^2}$ ($a \equiv |\vec{a}|$ for any
vector $\vec{a}$) with $W=\sqrt{M_d^2+ \vec{q}^{\,\,2}} + \omega$ as 
invariant mass of the photon deuteron system in the c.m.\ frame and $M_d$ as
deuteron mass. Finally, ${\widehat R}$ denotes the so-called dressing operator
describing hadronic off-shell effects~\cite{ScA01a,ScA01b,ScA01c}. 

The quantity ${\cal J}_{loop,\, sub}^{\mu}$ in (\ref{3_final})
denotes the e.m.\ corrections arising from the pion-nucleon loop which
have been discussed in detail in~\cite{ScA01c} and can be interpreted
as e.m.\ off-shell contributions. For the
transition to virtual photons e.m.\ form factors have to be
incorporated. Moreover, the propagators of the loop diagrams with an
intermediate $\gamma^{\ast} \pi N$-state have to be modified due to
the fact that $\omega \ne q$ in electrodisintegration. 

For the one-body ($\gamma + {\bar N} \rightarrow \Delta$)-transition
current, we take the usual nonrelativistic form including only the
dominant $M1$-transition 
\begin{eqnarray}
\bra{\vec{p}_{\Delta}} 
 \vec{\jmath}_{\Delta{\bar N}}(W_{sub}, \omega,\vec{q}\,)
 \ket{\vec{p}_{\bar N}}
&=& \delta(\vec{p}_{\Delta}-\vec{p}_{\bar N}-\vec{k})\,
 \frac{e\,\tau_{\Delta {\bar N},\,0}}{2M_N}
{\widetilde G}_{M1}^{\Delta {\bar N}}(W_{sub} + i \epsilon,
\omega, \vec{q}\,)
 \,i\vec{\sigma}_{\Delta {\bar N}}\times\vec{q}_{\gamma N}\,,
\label{ndeltacurrent}
\end{eqnarray}
where
\begin{eqnarray}\label{para_delta1}
\vec{q}_{\gamma N}&=&\vec{q} - 
\frac{M^{res}_{\Delta} - M_N}{M^{res}_{\Delta}}\,\vec{p}_{\Delta}\,
 \quad \mbox{with} \quad M^{res}_{\Delta} = 1232 \, \mbox{MeV}\,.
\end{eqnarray}
As has been outlined in detail in~\cite{ScA01b}, the e.m.\
coupling constant ${\widetilde G}_{M1}^{\Delta {\bar N}}$ 
has been fixed for real photons by a fit of the
$M_{1+}^{(3/2)}$-multipole of pion photoproduction on the nucleon. Due
to nonresonant rescattering mechanisms, it becomes complex
and dependent on the invariant mass $W_{sub}$ of the $\pi N$-subsystem
for which we adopt the spectator-on-shell approach~\cite{ScA01b}. 
Moreover, when embedded into a nuclear medium, ${\widetilde
G}_{M1}^{\Delta {\bar N}}$ depends also on the photon momentum $\vec{q}$
and the photon energy $\omega$. As has been shown in~\cite{Sch99}, 
at least for photodisintegration a so-called on-shell prescription
\begin{equation}\label{on-shell}
 {\widetilde G}_{M1}^{\Delta {\bar N}}(W_{sub}+i \epsilon,\omega, \vec{q}\,) 
 \rightarrow 
 {\widetilde G}_{M1}^{\Delta {\bar N}}(W_{sub}+ i \epsilon) 
 \end{equation} 
turns out to be very accurate, simplifying considerably the numerical
evaluation. The resulting coupling constant used in this paper
is parametrized as follows
\begin{equation}\label{delta-coupling}
   \widetilde{G}_{M1}^{\Delta {\bar N}}(z=W_{sub}+i \epsilon)=
  \widetilde{\mu}_{\Delta N}(W_{sub})\,e^{i\widetilde{\Phi}(W_{sub})}
\end{equation}
with
\begin{eqnarray}\label{para_delta2}
 \widetilde{\mu}_{\Delta N}(W_{sub})=
 \mu_0+\mu_2\left(\frac{q_{\pi}}{m_{\pi}}\right)^2
    +\mu_4\left(\frac{q_{\pi}}{m_{\pi}}\right)^4\,\mbox{ and }
     \,\, \widetilde{\Phi}(W_{sub})=\frac{q_{\pi}^3}{a_1+a_2 q_{\pi} +
  a_3 q_{\pi}^2 +a_4 q_{\pi}^3}\,,
\end{eqnarray}
where the on-shell pion momentum $q_{\pi}$ is a function of $W_{sub}$ 
according to $W_{sub}=M_N+\frac{q^2_{\pi}}{2 M_N}+\sqrt{m^2_{\pi}+q^2_{\pi}}$, 
while for $W_{sub} < M_N + m_{\pi}$ 
we have set $q_{\pi}=0$. The parameters in (\ref{para_delta2})
have been fitted to the resulting coupling constant 
``$\widetilde{G}_{M1}^{\Delta {\bar N}}(\mbox{eff1})$" discussed
in~\cite{ScA01b} and are listed in Table~\ref{tab1}. In addition a 
contribution from the $\gamma N\Delta$-transtion charge density is 
included. 
For virtual photons, again $\widetilde{G}_{M1}^{\Delta {\bar N}}$ must
be multiplied with an appropriate e.m.\ transition form factor which
for simplicity has been chosen in the dipole form. We are aware of the fact
that with increasing momentum transfer one finds a slightly stronger 
fall-off of $\widetilde{G}_{M1}^{\Delta {\bar N}}(Q^2)$ with 
$Q^2=q^2-\omega^2$ compared to the dipole form~\cite{TiD03}. 
However, we restrict ourselves here to kinematics with 
$Q^2 \leq 4$ fm$^{-2}$ where the difference does not matter.

With respect to the two-body currents, we consider, as in~\cite{ScA01b}, 
meson retardation in the pure pion exchange contributions whereas
retardation in $\gamma \pi\rho /\omega$-MEC turned out to be 
unimportant in photodisintegration~\cite{Sch99}, but very CPU-time consuming.
Therefore, in order to facilitate our numerical evaluation, we have
neglected retardation in the latter contribution. Furthermore, we include in
addition static $\rho$-exchange as well as various $\Delta$-MEC which
have been discussed in~\cite{ScA01b}. Similar to the e.m.\ loop
corrections the transition from photo- to electrodisintegration leads
to some modifications in the intermediate $NN \pi
\gamma^{\ast}$-propagators as well as for the incorporation of an
e.m.\ form factor. 

In the static model the e.m.\ loop contributions ${\cal J}_{loop,\,
sub}^{\mu}$ to the nucleonic one-body current
are neglected naturally and the hadronic dressing factor ${\widehat
R}$ is replaced by 1. Obviously, the MEC contain only the usual static
terms. Note that in the static limit the recoil contributions do not
contribute due to their cancellation against the wave function
renormalization~\cite{GaH76}. Moreover, in the static limit there are
no nonrelativistic MEC-contributions to the charge operator.

For the numerical evaluation of the $T$-matrix for deuteron 
electrodisintegration, we use the standard multipole decomposition of
the e.m.\ current. Moreover, we take advantage of Siegert's theorem
which allows to express the dominant contributions of the transverse
electric multipoles via the charge multipoles.

\section{Results and conclusions}
\label{result}
Neglecting polarization effects the differential cross section for
deuteron electrodisintegration in the one-photon exchange
approximation is determined by four structure functions, two diagonal
ones $f_L$ and $f_T$ and two interference ones $f_{LT}$ and
$f_{TT}$~\cite{FaA79}. They are functions of the squared three 
momentum transfer $q^2$ in the c.m.\ system, the final state
c.m.-energy $E_{np} = W - 2 M_N$, and the angle $\theta$ between $\vec
q$ and the relative neutron-proton momentum in the final
neutron-proton c.m.-system. 

\begin{figure}[h]
\centerline{\psfig{figure=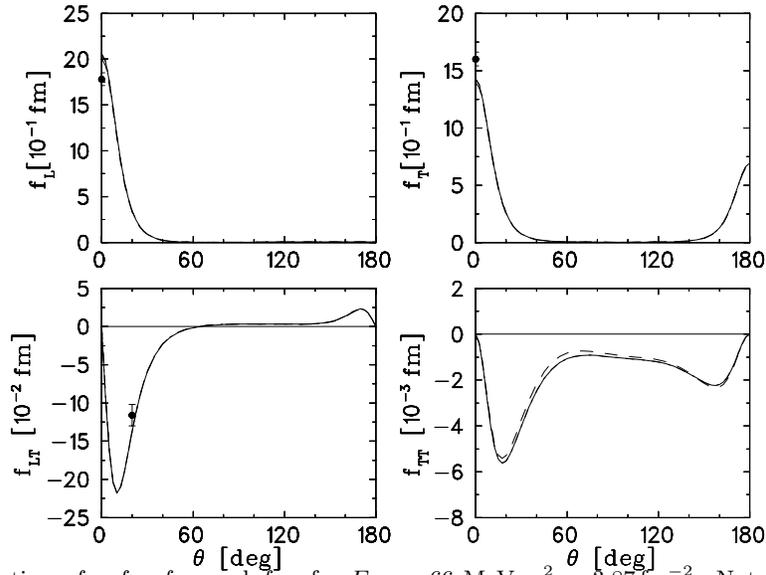,width=10cm,angle=0}}
\vspace*{-0.5cm}
\caption{The structure functions $f_L$, $f_T$, $f_{LT}$ and $f_{TT}$ for 
$E_{np} = 66$ MeV, $q^2= 3.87 \, \mbox{fm}^{-2}$.
Notation of the curves: dashed: static approach; full: retarded
approach. Experimental data from {\protect \cite{JoM96}}.} 
 \label{figure3}
\end{figure}

We begin the discussion with the experiment of Jordan et
al.~\cite{JoM96} for the kinematics $E_{np} = 66$ MeV, $q^2=
3.87$~fm$^{-2}$ which has been performed in order to extract the
structure functions $f_{L}$, $f_{T}$ and $f_{LT}$. Our predictions for
these structure functions using both static and retarded operators are
shown in Fig.~\ref{figure3} together with the few existing
experimental data points. Since this kinematics is well below pion
threshold it is not surprising that both approaches yield almost
identical results -- in fact only for the smallest structure function
$f_{TT}$ the corresponding curves can be distinguished --
demonstrating that both approaches are indeed equivalent in this
kinematic region. A similar result has also been found in
photodisintegration~\cite{ScA01b}. The agreement with experiment is
quite satisfactory but more data are certainly needed for a more
critical comparison. 

\begin{figure}
\centerline{\psfig{figure=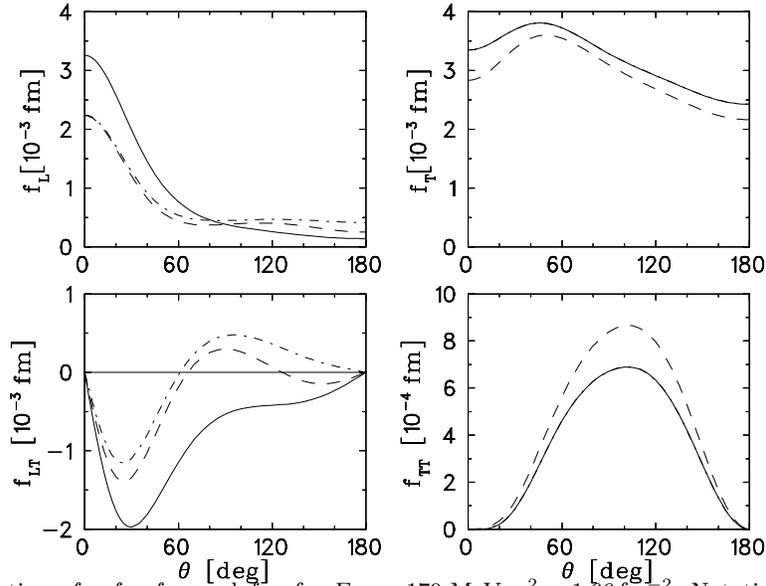,width=10cm,angle=0}}
\vspace*{-0.5cm}
\caption{
The structure functions $f_L$, $f_T$, $f_{LT}$ and $f_{TT}$ for 
$E_{np} = 179$ MeV, $q^2= 1.66 \, \mbox{fm}^{-2}$. Notation of the curves
as in Fig.~\ref{figure3}. The additional dash-dotted curves
represent the results of the retarded approach where the Coulomb monopole 
contribution of the recoil charge operator is switched off.}
\label{figure4}
\end{figure}

\begin{figure}
\centerline{\psfig{figure=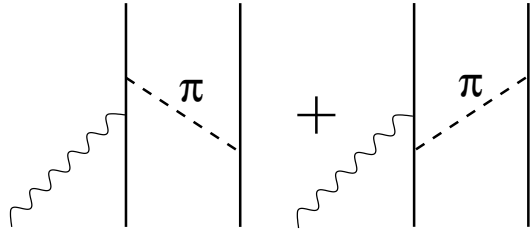,width=7cm,angle=0}}
\caption{
Diagrammatic representation of the recoil contributions to the
effective charge and current operators. The coupling of the photon
to the nucleon is given solely by the nonrelativistic charge and 
spin/convection current, respectively.}
\label{figure4a}
\end{figure}

\begin{figure}
\centerline{\psfig{figure=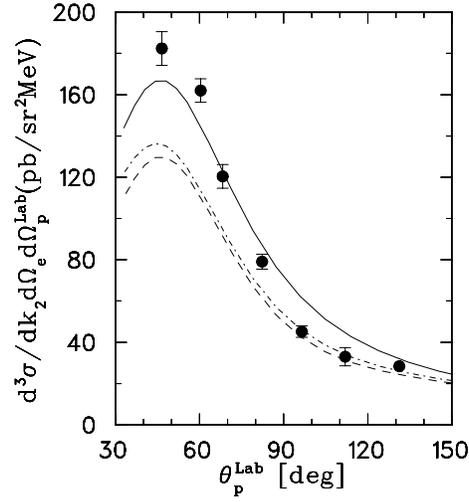,width=6cm,angle=0}}
\caption{The lab frame differential cross section for 
$E_{np} = 179$ MeV, $q^2= 1.66 \, \mbox{fm}^{-2}$ and an electron scattering
angle of $\theta_e^{Lab}=25^{\circ}$. The lab angle of the outgoing proton 
is denoted by $\theta_p^{\mathrm{Lab}}$. Experimental data from
{\protect \cite{TuB84}}.}
 \label{figure5}
\end{figure}

The situation changes completely if the excitation energy is above
pion threshold. As a first example, we consider in 
Fig.~\ref{figure4} the kinematics $E_{np}=179$ MeV, $q^2 =
1.66$~fm$^{-2}$ of an experiment by Turck-Chi\a`eze et
al.~\cite{TuB84}. Whereas $f_T$ is only moderately affected 
by retardation, the other structure functions, in particular 
$f_L$ and $f_{LT}$ are much more sensitive to the inclusion
of retardation. At forward angles $f_L$ is enhanced by
roughly 45 percent while above $\theta \approx 90^{\circ}$ it is
significantly lowered. Especially $f_{LT}$ increases strongly in
absolute magnitude. A detailed analysis has shown that this strong
retardation effect is mainly due to the charge recoil contribution 
to the Coulomb monopole (see Fig.~\ref{figure4a} for 
a graphical representation). This is demonstrated in
Fig.~\ref{figure4} by the dash-dotted curves for which this
contribution is switched off. The remaining retardation effect is
significantly smaller. 
A comparison with the experimental diffferential coincidence cross section
is shown in Fig.~\ref{figure5}. One readily notes a considerable 
enhancement of the cross section in the retarded approach improving
significantly the agreement with the data. But the
agreement is not perfect, the theory being a little too low in the
maximum around $\theta_p^{\mathrm{Lab}}=45^\circ$ and somewhat too 
large at higher angles. Again, the most important contribution of 
retardation is due to the recoil charge contribution to the monopole.

\begin{figure}
\centerline{\psfig{figure=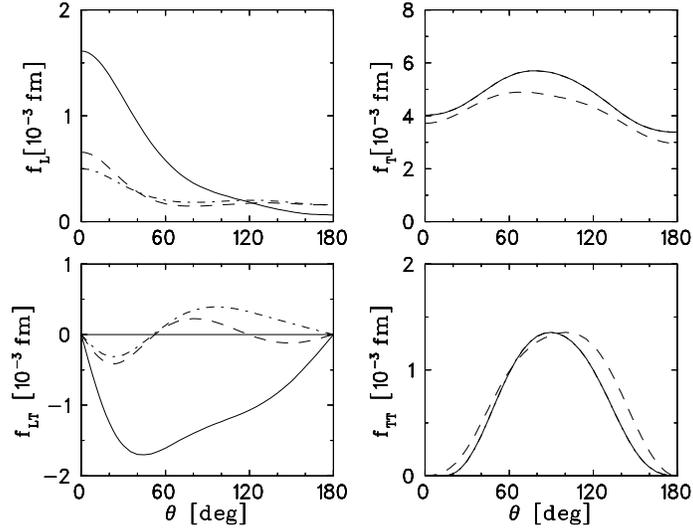,width=9cm,angle=0}}
\caption{The structure functions $f_L$, $f_T$, $f_{LT}$ and $f_{TT}$ for 
$E_{np} = 280$ MeV, $q^2= 2.47$~fm$^{-2}$. Notation of the curves
as in Fig.~\ref{figure4}.}
\label{figure6}
\end{figure}

\begin{figure}
\centerline{\psfig{figure=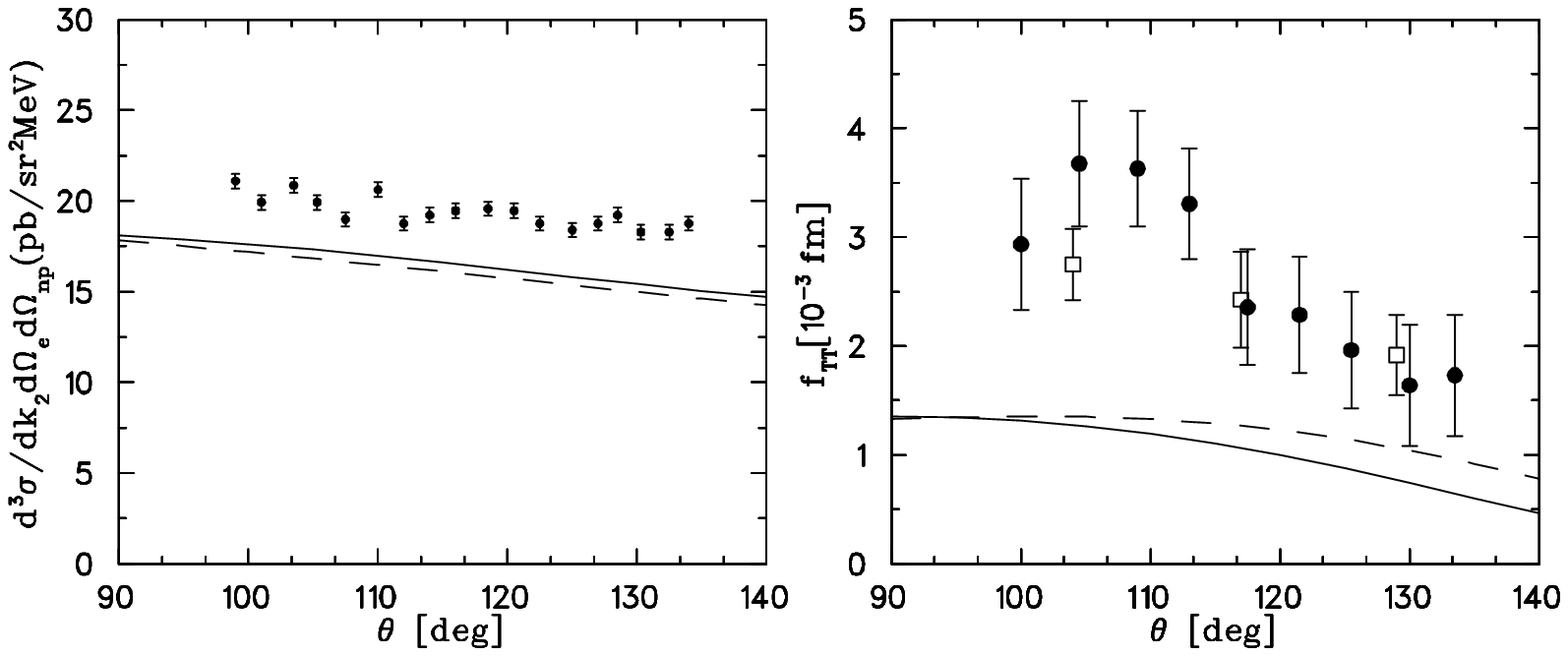,width=10cm,angle=0}}
\caption{The  differential cross section (left panel) 
and the $f_{TT}$-structure function (right panel)
for  $E_{np} = 280$ MeV, $q^2= 2.47$~fm$^{-2}, \theta_e=30^{\circ}$.
Notation as in Fig.~\ref{figure3}. Experimental data from 
{\protect \cite{PeA96}}.}
\label{figure7}
\end{figure}

The importance of retardation becomes even more striking for
excitation energies $E_{np}$ in the $\Delta$-region. This is
demonstrated in Figs.~\ref{figure6} and \ref{figure7} for the
kinematics of a more recent experiment by Pellegrino et
al.~\cite{PeA96} with $E_{np}=280$ MeV and  $q^2 =
2.47$~fm$^{-2}$. The effects of retardation on $f_{L}$ anf $f_{LT }$
in Fig.~\ref{figure6} are even stronger than in the previous
example. However, in the differential cross section the retardation
effects in the different structure functions appear to cancel each
other to a large extent so that static and retarded approaches yield
very close results in the measured angular range $95^{\circ} < \theta <
135^{\circ}$ (see Fig.~\ref{figure7}) but lead to a slight
underestimation of the experiment. The better agreement between 
theory and experiment reported in~\cite{PeA96} is obtained by the use
of a modified $\gamma N \Delta$-coupling which is considerably
stronger than ours given in Table~\ref{tab1} due to the assumption of a
vanishing nonresonant contribution to the
$M_{1+}^{(3/2)}$-multipole. However, as has been discussed
in~\cite{ScA01b}, this approach causes formal inconsistencies and is,
therefore, questionable. 

Comparing the static as well the retarded results for $f_{TT}$ in
Fig.~\ref{figure7} to experimental data, one notes a sizable
underestimation of both approaches. This discrepancy may have
different origins. First of all, one has to be aware of the fact that
the majority of the data (full circles in the right panel of 
Fig.~\ref{figure7}) has been extracted under the assumption $f_{LT} = 0$ 
which is well justified only in the static, but not all in the
retarded approach. Only the open square data points 
have been obtained without this assumption. Concerning the theoretical
uncertainties, the present approach does not contain all relativistic 
contributions of leading order in $p/M_N$ as is the case in the
approach, for example, of Ritz et al.~\cite{RiG97}. On the other hand,
among the relativistic contributions retardation effects are unique
with respect to the singularity structure of the corresponding
meson-nucleon propagators above pion threshold which cannot be treated
within a Taylor expansion in $p/M_N$. Therefore, concerning the role
of retardation, the approach of~\cite{RiG97} cannot be considered
realistic above pion threshold. However, the model of Ritz et al.\
is useful to estimate the role of boost effects (which do not contain
any singularities) which are still  missing in the present
approach. It turns out that for the kinematics of Fig.~\ref{figure7}
they are small. Moreover, we have checked 
that additional, not yet included mechanisms like static $\sigma$- 
or $\delta$-MEC, a nonvanishing $G_{En}$ or a possible $E2$-excitation 
of the $\Delta$ 
are small and cannot explain the noted discrepancy between theory and
experiment in Fig.~\ref{figure7}.

In summary, we have studied the role of retardation effects in
electrodisintegration of the deuteron for various kinematics. It
turns out that especially the recoil charge contribution to the
monopole, which is not present in static approaches, is very important
for excitation energies above pion threshold, leading to dramatic
changes in the structure functions $f_L$ and $f_{LT}$ whereas the
other structure functions $f_T$ and $f_{TT}$ are much less affected. 

With respect to future developments, one important step concerns an
improved treatment of the final-state-interaction. As has already been
noted in~\cite{ScA01a}, the description of the 
phase shifts in several NN-partial waves is only fairly well above 
pion threshold for our NN-interaction model. This touches a general
problem of realistic NN-interactions: In contrast to the energy regime
below pion threshold, at present no ``high precision" NN-interaction is
available for energies above pion threshold. However, this is an
indispensable ingredient for any theoretical description of e.m.\
reactions on the deuteron above pion threshold like photo- or
electrodisintegration, but also meson production on the deuteron. 
A further interesting topic will be the study of polarization observables
which supposedly are more sensitive to interaction effects.

\begin{table}[h]
\caption{Parameters of the $\gamma N\Delta$-coupling in (\ref{para_delta2}).}
\begin{ruledtabular}
\begin{tabular}{ccccccc}
$\mu_0$ & $\mu_2$ & $\mu_4$ & $a_1$~[fm$^{-3}$] & $a_2$~[fm$^{-2}$] & 
$a_3$~[fm$^{-1}$] & $a_4$ \\
\colrule
5.0912 &  0.017511 & $-0.017652$ & 0.5564 & 5.4317 & $-3.5987$ & 3.8491\\
\end{tabular}
\end{ruledtabular}
\label{tab1}
\end{table}

\begin{thebibliography}{99}
\bibitem{ScA98}
M.\ Schwamb, H.\ Arenh\"ovel, P.\ Wilhelm, Th.\ Wilbois,
Phys.\ Lett.\ B 420 (1998) 255.
\bibitem{ScA01a}
M.\ Schwamb, H.\ Arenh\"ovel, Nucl.\ Phys.\ A 690 (2001) 647.
\bibitem{ScA01b}
M.\ Schwamb, H.\ Arenh\"ovel, Nucl.\ Phys.\ A 690 (2001) 682.
\bibitem{ScA01c}
M.\ Schwamb, H.\ Arenh\"ovel, Nucl.\ Phys.\ A 696 (2001) 556.
\bibitem{Her99}
C.\ Herberg et al., Eur.\ Phys.\ J.\ A 5 (1999) 131.
\bibitem{Ost99}
M.\ Ostrick et al., Phys.\ Rev.\ Lett.\ 83 (1999) 276.
\bibitem{Pas99}
I.\ Passchier et al., Phys.\ Rev.\ Lett.\ 82 (1999) 4988.
\bibitem{ElF88}
Ch.\ Elster,  W.\ Ferchl\"ander, K.\ Holinde, D.\ Sch\"utte, 
R.\ Machleidt, 
 Phys.\ Rev.\  C 37 (1988) 1647.
\bibitem {GrS82}
A.\ M.\ Green, M.\ E.\ Saino, J.\ Phys.\ G 8 (1982) 1337.
\bibitem{MaH87a}
R.\ Machleidt, K.\  Holinde, Ch.\ Elster, Phys.\ Rep.\  149 (1987) 1.
\bibitem{WiB93}
T.\ Wilbois, G.\ Beck, H.\ Arenh\"ovel, Few Body Sys.\ 15 (1993) 39.
\bibitem{Sch99}
M.\ Schwamb, PhD-Thesis, Mainz 1999.
\bibitem{TiD03}
L. Tiator, D. Drechsel, S.S. Kamalov, S.N. Yang, Eur. Phys. J. A 17 (2003) 357.
\bibitem{GaH76}
M.\ Gari, H.\ Hyuga, Z.\ Phys.\ A 277 (1976) 291.
\bibitem{FaA79}
W.\ Fabian, H.\ Arenh\"ovel, Nucl.\ Phys.\ A 314 (1979) 253.
\bibitem{JoM96}
D.\ Jordan et al.,  Phys.\ Rev.\ Lett.\ 76 (1996) 1579.
\bibitem{TuB84}
S.\ Turck-Chi\a`eze et al., Phys.\ Lett.\ B 142 (1984) 145.
\bibitem{PeA96}
A.\ Pellegrino et al., Phys.\  Rev.\ Lett.\ 78 (1997) 4011.
\bibitem{RiG97}
F.\ Ritz, H.\ G\"oller, T.\ Wilbois, H.\ Arenh\"ovel,
Phys.\ Rev.\ C 55 (1997) 2214.
\end{thebibliography}
\end{document}